\documentclass[12pt]{iopart}
\usepackage{iopams}
\usepackage{iopams_add}
\usepackage{cite}
\usepackage{color}
\usepackage{graphicx}
\begin{document}
\title[]{Magnetization and susceptibility of polydisperse ferrofluids}
\author{I. Szalai$^1$, S. Nagy$^2$, and S. Dietrich$^{3,4}$}

\address{$^1$Institute of Physics and Mechatronics, 
University of Pannonia, H-8201 Veszpr\'em, PO Box 158, Hungary\\
$^2$Institute of Mechanics and Mechatronics, The University of West Hungary, 
H-9401 Sopron, PO Box 132, Hungary\\
$^3$Max-Planck-Institut f\"ur Intelligente Systeme, Heisenbergstr. 3, D-70569 Stuttgart, Germany\\
$^4$IV. Institut f\"ur Theoretische Physik, Universit\"at Stuttgart, Pfaffenwaldring 57, 
D-70569 Stuttgart, Germany}
\eads{szalai@almos.vein.hu, nagy.sandor@fmk.nyme.hu, dietrich@is.mpg.de}
\begin{abstract}
On the basis of the mean spherical approximation of multicomponent dipolar hard sphere mixtures 
an analytical expression is proposed for the magnetic field dependence 
of the magnetization of size polydisperse ferrofluids. The polydispersity of the particle diameter is described by the  gamma  distribution function. 
Canonical ensemble Monte Carlo simulations have been 
performed in order to test these theoretical results for the initial 
susceptibility and the magnetization. The results for the magnetic properties 
of the polydisperse systems turn out to be in quantitative agreement with our present 
simulation data.
In addition, we find good agreement between our theory and experimental data 
for magnetite-based ferrofluids. 
\end{abstract}
\section{Introduction}
Magnetic fluids are colloidal suspensions of single domain ferromagnetic grains dispersed 
in a solvent \cite{rosensweig1}.
In order to keep such suspensions stable the grains have to be coated with polymers or surfactant layers or by using electric double layer formation in the case of water-based ferrofluids.
Accordingly the ferromagnetic particles typically come in different sizes, 
commonly ranging from 5 nm to 50 nm.
Each particle of a ferrofluid possesses a permanent magnetic dipole moment. 
Therefore, in many cases ferrofluids can be described as dipolar liquids.
Actual ferrofluids are more or less polydisperse. 
This means that the particles may have different sizes and thus carry different magnetic moments  
which are proportional to their volume. 
A polydisperse fluid can be considered as a mixture consisting of a 
large number of components, with an essentially continuous distribution of the particle size.\\ 
In dilute magnetic fluids the magnetic dipole-dipole interaction 
can be neglected as compared with the interaction between the particles and an external magnetic field. 
Therefore, in an applied field the magnetization of these 
systems can be described by the Langevin theory. 
At higher concentrations the effective interaction of disolved magnetic 
particles can be modeled in terms of dipolar hard spheres (DHS). 
Their interaction potential is the sum of 
the isotropic hard sphere (HS) and the anisotropic dipole-dipole interaction.

Over the last few decades theoretical descriptions have evolved which allow one to study the 
initial susceptibility and the magnetization of mono- and polydisperse magnetic colloids. 
Following Weiss' idea,  Pshenichnikov \cite{pshe1,pshe2} proposed the so-called 
modified mean-field model for calculating these quantities. Later, using the mean spherical 
approximation (MSA) results of Wertheim \cite{wertheim1}, Morozov and Lebedev \cite{morozov1} 
extended the applicability of MSA to the calculation of magnetizations of ferrofluids.
On the basis of thermodynamic perturbation theory, 
Ivanov and Kuznetsova \cite{ivanov1,ivanov2} proposed a more sophisticated theory for these calculations. 
In order to assess the reliability of the aforementioned theoretical methods 
Ivanov \textit{et al} \cite{ivanov3} compared them with Monte Carlo (MC) and 
molecular dynamics (MD) simulation data as well as with experimental data.
Along these lines Huke and L\"ucke \cite{huke1} proposed a cluster expansion approach 
for monodisperse systems, which later was extended to polydisperse systems 
\cite{huke2}.

Starting from Wertheim's \cite{wertheim1} analytical MSA results, 
within the framework of density functional theory (DFT) two of the present authors have also 
proposed \cite{szalai1} an equation for the 
magnetization of  \textit{mono}disperse ferrofluids, which turns out to be simpler 
than the corresponding equation of 
Morozov and Lebedev \cite{morozov1}. Nonetheless quantitative agreement was found between these DFT results and corresponding canonical MC simulation data. Based on the multi-component MSA solution 
obtained by Adelman and Deutch \cite{adelman1} this theoretical approach was extended to the description of the magnetization of \textit{multi}-component systems \cite{szalai2}. For the studied two- and three-component systems this theory turned out to be reliable as compared with the corresponding MC simulation data.
 
On the basis of a natural extension of the multi-component MSA to polydisperse systems, 
in Ref. \cite{szalai2} we proposed an equation for the magnetization of \textit{polydisperse} magnetic fluids. 
In the following, for various polydispere systems we compare our aforementioned theory with 
new corresponding MC simulations and experimental data.  
\section{Theory}

For ferromagnetic grains the dipole moment of a particle is given by
\begin{equation}
m(\sigma)=\frac{\pi}{6}M_0\sigma^3,
\label{eq:mom}
\end{equation}
where $M_0$ is the bulk saturation magnetization of the core material 
(for magnetite $M_0=480$ kA/m at room temperature) and $\sigma$ is the diameter of the particle. 
The Langevin susceptibility of a polydisperse system \cite{ivanov3,szalai2} is
\begin{equation}
\chi_L=\frac{1}{3}\beta\rho\int_0^{\infty}d{\sigma}p(\sigma)m^2(\sigma),
\label{eq:slan}
\end{equation}
where $\rho=N/V$ is the number density in the volume $V$ of the system, 
$\beta=1/(k_BT)$ is the inverse thermal energy with the Boltzmann constant $k_B$ 
and temperature $T$, and $p(\sigma)$ is the probability distribution for the magnetic core diameter. 
(We note that the original multi-component MSA \cite{adelman1} is valid for equally 
sized hard spheres; however, it was first formally extended to polydisperse systems 
by Morozov and Lebedev \cite{morozov1} and by Ivanov \textit{et al.} \cite{ivanov3}.)
The dependence of the magnetization $M$ of the polydisperse system on an external 
magnetic field $H$ is given by an implicit equation \cite{szalai2}: 
\begin{equation}
M=\rho\int_0^{\infty}d{\sigma}p(\sigma)m(\sigma)
L\left[{\beta}m(\sigma)\left(H+\frac{(1-q(-\xi))}{\chi_L}M\right)\right],
\label{eq:magn}
\end{equation}
where $L(z)=\coth(z)-1/z$ is the Langevin function, and $\xi$ is the implicit solution of the corresponding 
MSA equation
\begin{equation}
 4\pi\chi_L=q(2\xi)-q(-\xi).
\label{eq:msaeq}
\end{equation}
In Eqs. (\ref{eq:magn}) and (\ref{eq:msaeq}) the function $q(x)$ is the reduced inverse compressibility 
function of hard spheres within the Percus-Yevick approximation:
\begin{equation}
q(x)=\frac{(1+2x)^2}{(1-x)^4}.
\end{equation}
According to Eq. (\ref{eq:magn}) the zero-field (initial) magnetic susceptibility 
of polydisperse system is
\begin{equation}
\chi_0=\frac{\chi_L}{q(-\xi)}.
\label{chi0}
\end{equation}
We note that for monodisperse systems (i.e., for $p(\sigma)=\delta(\sigma-\sigma_m)$, where $\sigma_m$ is the 
particle diameter of monodisperse grains and $\delta$ is Dirac's delta function) 
Eq. (\ref{eq:magn}) yields our previous result 
\cite{szalai1}, while Eqs. (\ref{eq:slan}) and (\ref{eq:msaeq}) render the results of 
Ref. \cite{wertheim1}.

The particle polydispersity of magnetic fluids is commonly described 
\cite{morozov1,ivanov1,ivanov2,ivanov3} by the gamma distribution 
\begin{equation}
p_a(\sigma;\sigma_0)=\frac{1}{{\sigma_0}}\left(\frac{\sigma}{\sigma_0}
\right)^a\frac{\e^{-\sigma/{\sigma_0}}}{\Gamma(a+1)}
\label{eq:dis}
\end{equation}
where $\sigma$ is the magnetic core diameter of the particles, $\Gamma(s)$ is the gamma function, 
and $a$ and $\sigma_0$ are the shape and the scale parameter of the distribution, respectively. In the following we provide expressions for certain quantities, which are important for characterizing the particle parameters and the thermodynamic state of polydisperse systems. 
The mean value of the particle diameter is
\begin{equation}
\langle{\sigma}\rangle_p=\int_0^{\infty}d\sigma\,p_a(\sigma;\sigma_0)\,\sigma=\sigma_{0}(a+1).
\label{meanv} 
\end{equation} 
The expression for the reduced density and for the mean value of the dipole moment requires 
to know the third moment of $p(\sigma)$:
\begin{equation}
\langle{\sigma^3}\rangle_p=\int_0^{\infty}d\sigma\,
p_a(\sigma;\sigma_0)\sigma^3={\sigma_0^3}\prod_{i=1}^3(a+i).
\end{equation}
For the average of the magnetic dipole moment of particles, from Eqs. (\ref{eq:mom}) and 
(\ref{eq:dis}) one finds
\begin{equation}
\!\!\!\!\!\!\!\!\!\!\!\!
\langle{m}\rangle_p=\int_0^{\infty}d\sigma\,p_a(\sigma;\sigma_0)m(\sigma)=
\frac{\pi}{6}M_0\langle{\sigma^3}\rangle_p=
\frac{\pi}{6}M_0{\sigma_0^3}\prod_{i=1}^3(a+i).
\label{mave}
\end{equation}
According to Eq. (\ref{eq:slan}) the Langevin susceptibility is proportional to the 
mean-square dipole moment:
\begin{eqnarray}
\langle{m^2}\rangle_p=\int_0^{\infty}d\sigma\,p_a(\sigma;\sigma_0)m^2(\sigma)=\nonumber\\
\left(\frac{\pi}{6}M_0\right)^2
\int_0^{\infty}d\sigma\,p_a(\sigma;\sigma_0)\sigma^6=
\left(\frac{\pi}{6}M_0\right)^2{\sigma_0^6}
\prod_{i=1}^6(a+i)
\end{eqnarray}
which contains the sixth moment of the diameter $\sigma$. The reduced mean-square 
dipole moment is defined as
\begin{equation}
\langle{m^{2}}\rangle_p^*=\frac{\langle{m^{2}}\rangle_p}{\langle{\sigma^3}\rangle_p{k_BT}}.
\label{dip2}
\end{equation}

In a $k$ component mixture of particles with different diameters $\sigma_i$ the number density 
of the system is characterized by the packing fraction
\begin{equation}
\eta=\frac{1}{V}\sum_{i=1}^{k}N_iv_i=\rho\sum_{i=1}^{k}\left(\frac{N_i}{N}\right)v_i=
\frac{\pi}{6}\rho\sum_{i=1}^{k}\left(\frac{N_i}{N}\right)\sigma_i^3,
\label{eq:etam}
\end{equation}
where $N_i$ is the number of particles of the $i$th component, 
$v_i$ is the volume of the $i$th type of particles, 
and $N=\sum_{i=1}^{k}{N_i}$ is the total number of particles. 
The natural generalization of Eq. (\ref{eq:etam}) to a continouos distribution 
system is
\begin{equation}
\eta=\frac{\pi}{6}\rho\int_0^{\infty}d\sigma\,p_a(\sigma;\sigma_0)\sigma^3
=\frac{\pi}{6}\rho\langle{\sigma^3}\rangle_p=
\frac{\pi}{6}\rho{\sigma_0^3}\prod_{i=1}^3(a+i).
\label{eq:etap}
\end{equation}
In Refs. \cite{szalai1,szalai2} the dependence of the magnetization on the magnetic field has been investigated at fixed reduced 
densities and dipole moments. By alluding to Eq. (\ref{eq:etap}), in the case of polydisperse fluids a reduced density $\rho^*$ can be defined as  
\begin{equation}
\rho^*=\rho\langle{\sigma^3}\rangle_p=\rho{\sigma_0^3}(a+1)(a+2)(a+3).
\label{den}
\end{equation}
This implies that $\eta=\pi\rho^*/6$. 
On the basis of Eqs. (\ref{meanv}) and (\ref{den}) in the limit ${a}\gg{1}$ the reduced density reduces to
\begin{equation}
\!\!\!\!\!\!\!\!\!\!\!\!\!\!\!
\rho^*=\rho\langle{\sigma}\rangle_p^3
\frac{(a+1)(a+2)(a+3)}{(a+1)^3}\,\,\,\,\,\,{\xrightarrow[{a}\gg{1}]{}}\,\,\,\,\,\,
\rho\langle{\sigma}\rangle_p^3=:\rho\sigma_m^3=:\rho_m^* \,.
\end{equation}
This relation defines the reduced density $\rho^*_m$ of the monodisperse system. 
Equations (\ref{eq:slan}), (\ref{dip2}), and (\ref{den}) lead to an equation for the Langevin susceptibility, expressed in terms of reduced variables:
\begin{equation}
\chi_L=\frac{1}{3}{\rho^*}\langle{m^{2}}\rangle_p^*\,\,.
\end{equation}

The comparison between theory and the simulation data can be carried out best by applying 
a scaled one-parameter probability distribution function (see, c.f., Eq (\ref{eq:reddis})). 
In view of the simulations this distribution function has to be discretized (see below).


\section{Monte Carlo simulations}
The reliability of these theoretical predictions has been assessed by performing NVT (canonical) 
Monte Carlo (MC) simulations. In the MC simulations we have discretized the polydisperse system 
by considering a system of $k$ components in which the particles belonging to different components differ with respect to their diameters and dipole moments.
The system is characterized by the following dipolar hard sphere (DHS) potential:
\begin{equation}
u_{ij}^{DHS}({\mathbf{r}}_{ij},\omega_i,\omega_j)= u_{ij}^{HS}(r_{ij})+
u_{ij}^{DD}({\mathbf{r}}_{ij},\omega_i,\omega_j),
\label{eq:dhs}
\end{equation}
where $u_{ij}^{HS}$ and $u_{ij}^{DD}$ are the hard-sphere and the dipole-dipole interaction 
potential, respectively. 
The hard sphere pair potential is given by
\begin{equation}
u^{HS}_{ij}({r}_{ij})=\left\{
        \begin{array}{lll}
     \infty &, & r_{ij} < (\sigma_{i}+\sigma_{j})/2 \\
      0 &, & r_{ij}\geq (\sigma_{i}+\sigma_{j})/2\,\,,
        \end{array} 
        \right. \
\label{eq:hs}
\end{equation}
where $\sigma_l$ is the diameter of the $l$th particle.
The dipole-dipole pair potential is
\begin{equation}
u^{DD}_{ij}({\mathbf{r}}_{ij},\omega_i,\omega_j)=-\frac{m_im_j}{r_{ij}^3}D(\omega_{ij},\omega_i,\omega_j),
\end{equation}
with the rotationally invariant function
\begin{eqnarray}
D(\omega_{ij},\omega_i,\omega_j)=\nonumber\\
3(\widehat{\mathbf{m}}_i(\omega_i)\cdot\widehat{\mathbf{r}}_{ij}(\omega_{ij}))
(\widehat{\mathbf{m}}_j(\omega_j)\cdot\widehat{\mathbf{r}}_{ij}(\omega_{ij}))-
(\widehat{\mathbf{m}}_i(\omega_i)\cdot\widehat{\mathbf{m}}_{j}(\omega_{j})),
\end{eqnarray}
where particle $i$ ($j$) located at ${\mathbf{r}}_i$ (${\mathbf{r}}_j$) has a diameter $\sigma_i$ ($\sigma_j$) and carries a dipole moment of strength 
$m_i={\pi}M_0\sigma_i^3/6$ ($m_j={\pi}M_0\sigma_j^3/6$) with an orientation of the dipole 
given by the unit vector $\widehat{\mathbf{m}}_i$ ($\widehat{\mathbf{m}}_j$)
with polar angles $\omega_i=(\theta_i,\phi_i)$ ($\omega_j=(\theta_j,\phi_j)$);
${\mathbf{r}}_{ij}={\mathbf{r}}_i-{\mathbf{r}}_j$ is the difference vector between the center of particle $i$ and the center of particle $j$ with $r_{ij}=|{\mathbf{r}}_{ij}|$. 
We have applied a homogeneous external magnetic field $\mathbf{H}$ to the system, the direction of 
which is taken to coincide with the direction of the $z$ axis. For a single magnetic dipole 
the magnetic field gives rise to the following additional contribution to the interaction potential:
\begin{equation}
u_i(\theta_i)=-\mathbf{m}_i\mathbf{H}=-m_iH\cos\theta_i,
\end{equation}
where the angle $\theta_i$ measures the orientation of the $i$th dipole relative to the field direction, i.e., the $z$ axis.

For the simulations we have used the discretized particle diameter distribution function of 
polydisperse system introduced in Ref. \cite{kristof1}. To this end, by using 
Eq. (\ref{meanv}) the probability distribution in Eq. (\ref{eq:dis}) can be rewritten as
\begin{equation}
p_a(\sigma;\sigma_0)=
\frac{1}{\langle\sigma\rangle_p}\frac{(a+1)^{a+1}}{\Gamma(a+1)}\left({\frac{\sigma}
{\langle\sigma\rangle_p}}\right)^a
\e^{-(a+1){\sigma}/{\langle\sigma\rangle_p}}. 
\label{eq:reddis}
\end{equation}
If one keeps $\langle{\sigma}\rangle_p\equiv\langle{\sigma}\rangle$ fixed, i.e., independent of 
$a$, this creates a new distribution
\begin{equation}
t_a(\sigma;\langle\sigma\rangle)=
\frac{1}{\langle\sigma\rangle}\frac{(a+1)^{a+1}}{\Gamma(a+1)}\left({\frac{\sigma}
{\langle\sigma\rangle}}\right)^a
\e^{-(a+1){\sigma}/{\langle\sigma\rangle}}
\label{eq:reddis2}
\end{equation}
with the properties
\begin{equation}
\langle{\sigma}\rangle_t=\langle{\sigma}\rangle,\,\,\,\,\,
\langle{\sigma^2}\rangle_t=\frac{a+2}{a+1}\langle{\sigma}\rangle^2,\,\,\,\,\,
\langle{\sigma^3}\rangle_t=\frac{(a+2)(a+3)}{(a+1)^2}\langle{\sigma}\rangle^3
\label{moms}
\end{equation}
so that the variance 
$\sqrt{\langle{\sigma^2}\rangle_t-\langle{\sigma}\rangle^2_t}=\langle{\sigma}\rangle/\sqrt{a+1}$ 
vanishes in the limit $a\rightarrow{\infty}$, in accordance with
\begin{equation}
\lim_{a\rightarrow\infty}t_a(\sigma;\langle{\sigma}\rangle)=
\delta(\sigma-\langle{\sigma}\rangle),
\end{equation}
which corresponds to the monodisperse limit. Equation (\ref{eq:reddis2}) shows that 
$\langle{\sigma}\rangle{t_a}(\sigma;\langle{\sigma}\rangle)$ as function of $\sigma/\langle{\sigma}\rangle$ depends on $a$ only. 
This form facilitates the representation of the aforementioned discretization, which is based on 
a fixed number of discrete fractions, containing $N_i$ particles with diameter $\sigma_i$. 
The number of particles in each fraction is given by the discretized distribution. 
We limit the number of fractions by requiring that there are at least 3 particles in each fraction.

By carrying out zero-field NVT ensemble simulations the initial (zero-field) magnetic susceptibility has been obtained from the fluctuations of the total magnetic dipole 
moment of the system (see Eq. (\ref{chi0})):
\begin{equation}
\chi_0=\frac{\beta}{3V}\left(\left\langle{\mathcal{M}^2}\right\rangle-
\left\langle{\mathcal{M}}\right\rangle^2\right),
\end{equation}
where $\mathcal{M}=\sum_{i=1}^N\mathbf{m}_i$ is the instantaneous magnetic dipole 
moment of the system, and the brackets denote the ensemble average.
In our simulations with an applied field the equilibrium magnetization of the polydisperse system is 
obtained from the equation
\begin{equation}
\mathbf{M}=\frac{1}{V}\left\langle{\mathcal{M}}\right\rangle.
\end{equation}
The long-ranged dipolar interactions have been treated using the reaction-field 
method with conducting boundary condition \cite{allen1}. 
In this case the applied external field is identical 
to the internal field acting on the particles throughout the simulation box. 
We have started our simulations from a spatially and orientationally disordered 
(randomly generated) initial configuration. 
Ahead of the production cycles, $0.5$ million equilibration cycles have been used. 
We have performed 
2-3 million production cycles within every simulation run, using $N=512$ particles. 
Estimates for the error bars have been obtained by 
dividing each run as a whole into 10-20 blocks and by 
calculating the standard deviation of the block averages. 

\begin{figure}
\begin{center} 
\scalebox{0.55}{*\includegraphics{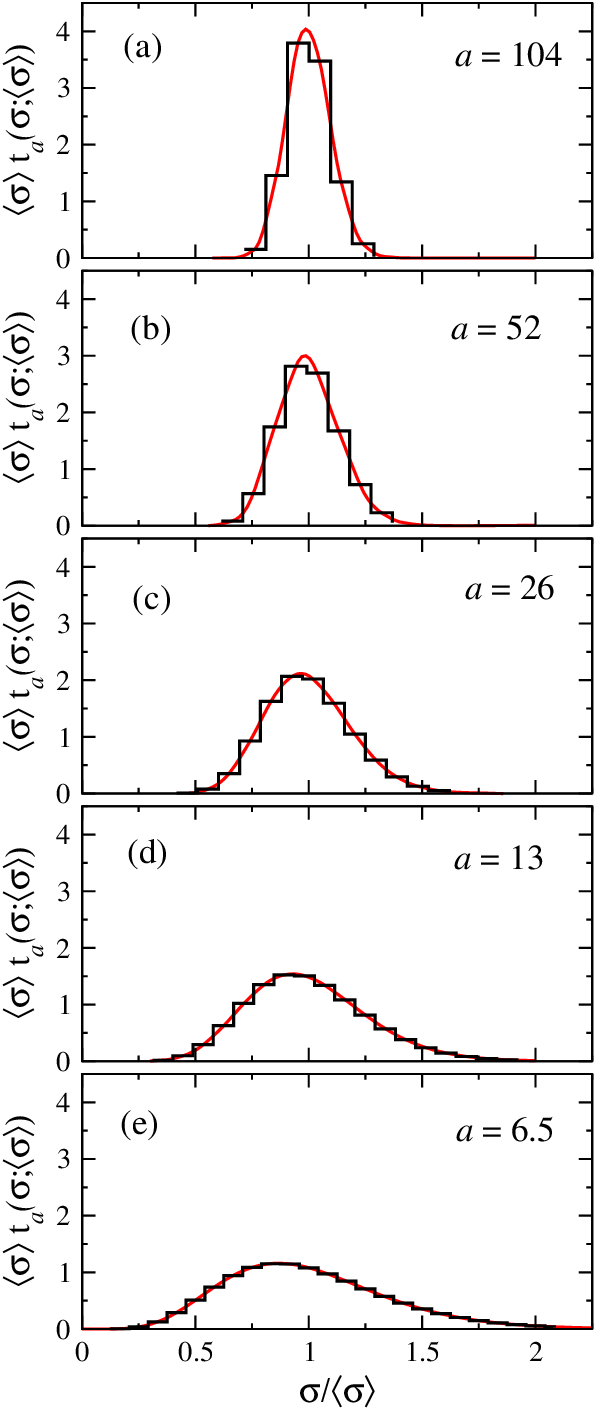}}
\end{center}
\caption{Discretization of the particle diameter distribution functions for various 
polydispersity shape parameter values $a$; $\langle{\sigma}\rangle$ is the mean 
value of the particle diameter. Red lines represent the continous gamma 
distribution functions (Eq. (\ref{eq:reddis2})), while the black lines represent the 
corresponding discretized particle distribution functions.}
\label{fig1}
\end{figure} 
\section{Numerical results and discussion}
First, we display the discretized particle diameter distribution functions for which the simulations have been carried out. 
Figures 1(a)-(e) show the rescaled discretized gamma distribution curves in comparison with 
the corresponding continuous ones for the values $a=104,\,52,\,26,\,13$, and $6.5$ of the shape parameter. 
One sees that the width of the distributions increases with decreasing values of the parameter $a$. 
The position $\sigma_{max}/\langle{\sigma}\rangle$ of the maximum of the 
distributions $\langle{\sigma}\rangle{t_a}(\sigma;\langle{\sigma}\rangle)$ tends to unity 
with increasing $a$, because $\sigma_{max}/\langle\sigma\rangle=a/(a+1)$. 
According to Figs. 1(a)-(e), for the various values of $a$ the number 
$N_a$ of particle fractions are $N_{104}=6$, $N_{52}=8$, $N_{26}=12$, $N_{13}=18$, and $N_{6.5}=24$. 
Also in their discretized version the distribution functions are normalized.
\begin{figure}
\begin{center} 
\scalebox{0.45}{*\includegraphics{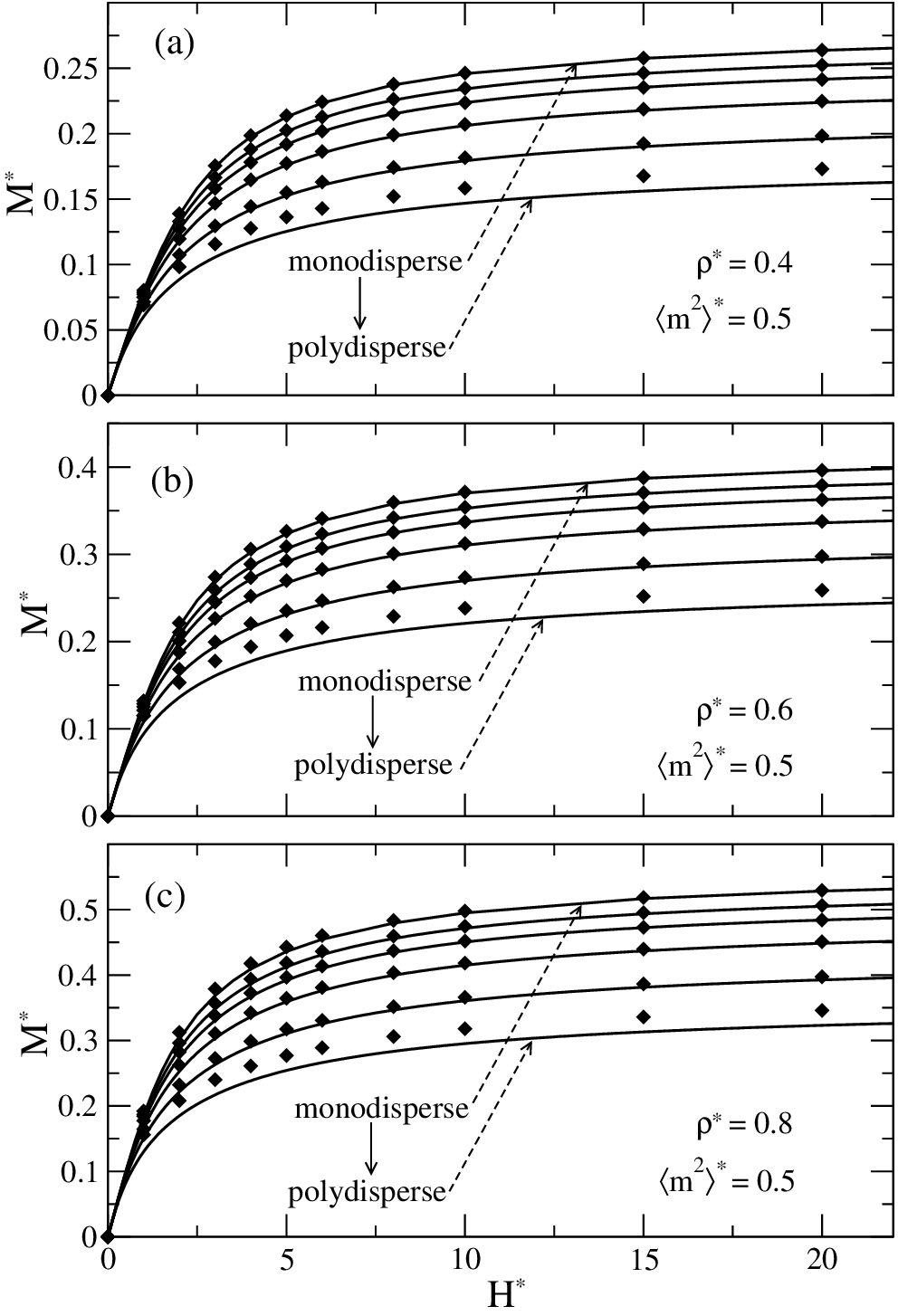}}
\end{center}
\caption{Magnetization curves $M^*=M\sqrt{\langle\sigma^3\rangle_p/(k_BT)}$ as function 
of $H^*=H\sqrt{\langle\sigma^3\rangle_p/(k_BT)}$ for polydisperse DHS fluids 
for six values of the shape parameter $a$ of the size distribution (Eq. (\ref{eq:dis})) 
(top to bottom for each density: $a=\infty$ (monodisperse), $a=104$, $a=52$, $a=26$, 
$a=13$, and $a=6.5$) for three different reduced densities $\rho^*$ (Eq. (\ref{den})) 
and the mean-square dipole moment $\langle{m^2}\rangle^*=0.5$ (Eq. (\ref{dip2})). 
According to Eq. (\ref{den}) changing $a$ with keeping $\rho^*$ fixed can be accomplished by 
varying $\sigma_0$.
For a given mean-square dipole moment and 
for a fixed reduced density the magnetization curves are shifted downwards upon 
enhancing polydispersity $a$. The data points correspond to MC data whereas 
the lines correspond to Eq. (\ref{eq:magn}).}
\label{fig2}
\end{figure} 
Figure 2 displays the magnetization curves of polydisperse DHS fluids 
with a mean-square dipole moment $\langle{m^2}\rangle^*=0.5$ (see Eq. (\ref{dip2})) 
for three values of the reduced number density $\rho^*$ (see Eqs. (\ref{den}) and (\ref{moms})) 
and for six values of the polydispersity shape parameter $a$ 
for each reduced density.
For a given reduced density the magnetization curves 
are shifted downwards upon decreasing $a$.
For the calculation of the magnetization curves the 
reduced density $\rho^*$ of the system has to be fixed. 
For a constant polydispersity $a$, this means that 
in a cubical simulation box of given volume $V=l^3$ and for a given number $N$ of 
particles the diameter scale $\sigma_0$ has to be fixed, 
due to Eq. (\ref{den}): 
\begin{equation}
\rho^*=\frac{N}{V}{\sigma_0^3}(a+1)(a+2)(a+3)=
N\left(\frac{\sigma_0}{l}\right)^3(a+1)(a+2)(a+3).
\end{equation}
In our simulations we have set simultaneously the reduced number density and the mean-square 
dipole moment, while we have changed the polydispersity (i.e., $a$) of the system. 
This is possible only if one changes the saturation magnetization parameter $M_0$ of the particles. 
Therefore the magnetization curves associated with different 
parameters $a$ (with $\rho^*$ and $\langle{m^2}\rangle^*$ fixed) correspond to distinct 
kinds of materials.
For strong magnetic fields $H^*$, we have found excellent quantitative agreement for all 
densities and polydispersities between the DFT 
results (see Eq. (\ref{eq:magn})) and the MC data. 
Close to the elbow of the magnetization curves the level of quantitative agreement 
is reduced, in particular for higher densities and polydispersities. We note that 
for equally sized DHS mixtures this range is also the most sensitive one concerning the 
agreement between theoretical results and MC simulation data \cite{szalai2}. 
Figure 3 displays the magnetization curves of polydisperse DHS fluids 
with mean-square dipole moment $\langle{m^2}\rangle^*=1$ for three values 
of the reduced density and for six values of the polydispersity shape parameter $a$. 
As in the case $\langle{m^2}\rangle^*=0.5$, for a given reduced 
density the magnetization curves 
are shifted downwards upon decreasing $a$. 
\begin{figure}
\begin{center} 
\scalebox{0.45}{*\includegraphics{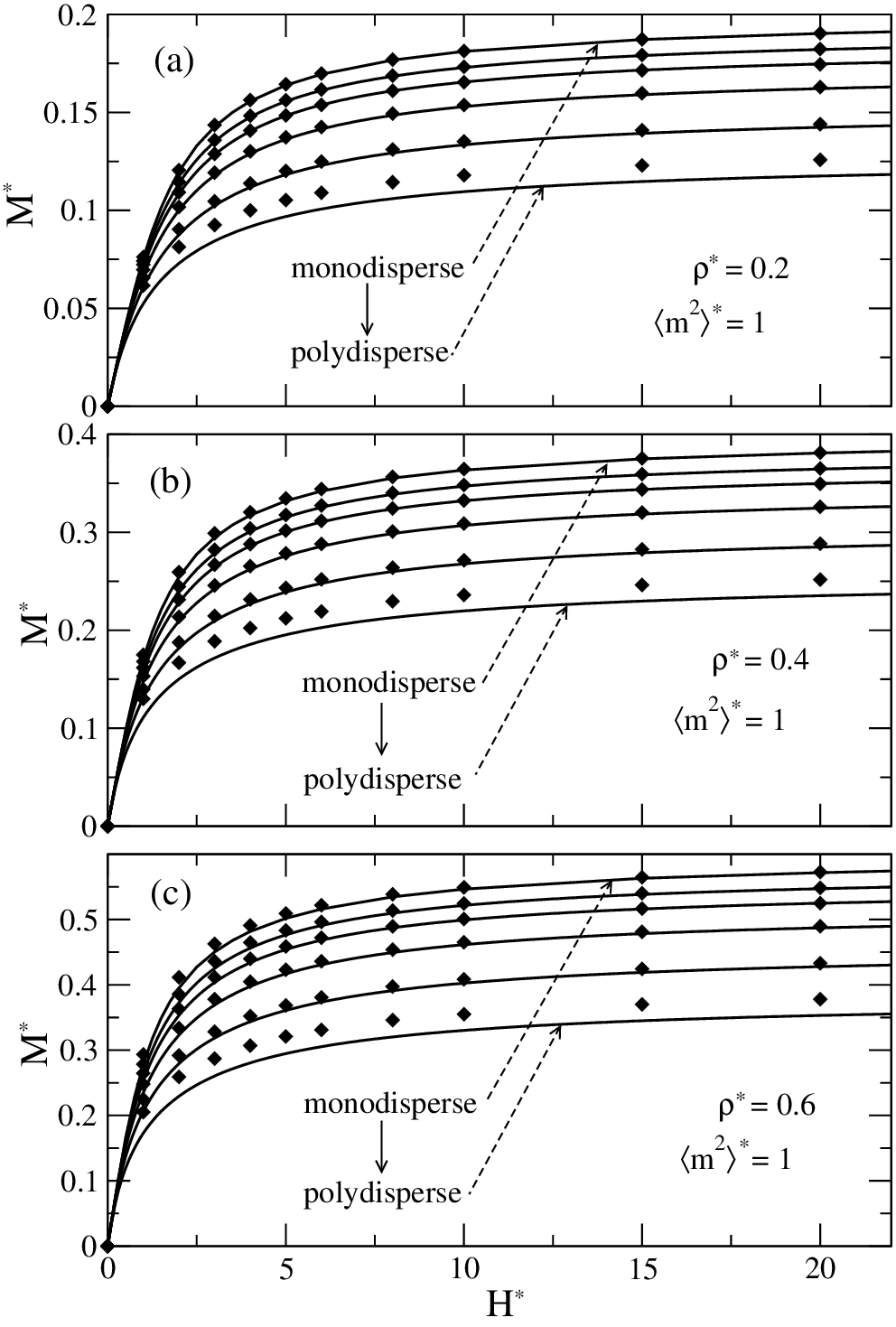}}
\end{center}
\caption{Same as Fig. 2 for $\langle{m^2}\rangle^*=1$ and $\rho^*=$ 0.2, 0.4, and 0.6.}
\label{fig3}
\end{figure} 
From the comparison between Fig. 3 with Fig. 2 one can infer that the quantitative 
agreement between the theoretical magnetization curves and the 
MC simulation data deteriorates slightly upon increasing the mean-square dipole moment. 
In view of this quantitative agreement between the theoretical results and the simulation 
magnetization data we can conclude 
that the magnetization equation of state (see Eq. (\ref{eq:magn})) is reliable 
up to values of the Langevin susceptibility $4\pi\chi_L=\frac{4\pi}{3}\rho^*{\langle{m^2}\rangle}^*\lesssim
\frac{4\pi}{3}\times0.4\times1\simeq{1.7}$.
\begin{figure}
\begin{center}
\scalebox{0.45}{*\includegraphics{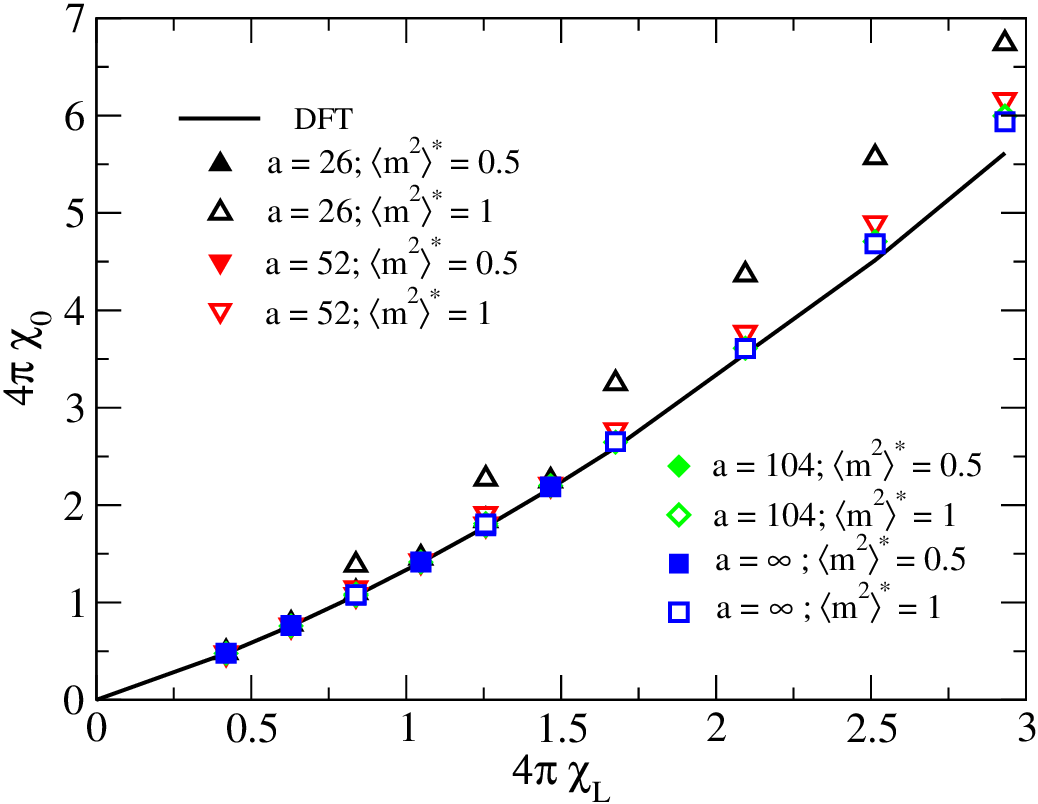}}
\end{center}
\caption{Zero-field susceptibility $\chi_0$ (Eq. (\ref{chi0})) of polydisperse 
systems as a function of the Langevin susceptibility $\chi_L$ (Eq. (\ref{eq:slan})). 
In terms of these quantities DFT (MSA) predicts the master curve given by 
the full line. The symbols represent the corresponding MC simulation data. Some of 
the symbols are not visible because they are lying on top of each other.}
\label{fig4}
\end{figure} 
Figure 4 shows the dependence of the zero-field susceptibility $\chi_0$ on the 
Langevin susceptibility $\chi_L$ for polydisperse DHS fluids as obtained from 
Eq. (\ref{chi0}) and from the numerical solution of Eq. (\ref{eq:msaeq}). 
One can see that the MSA based DFT (continuous line) provides a  master curve which is the same 
for various polydisperse systems. For low polydispersity ($a=\infty$, 104, and 52) 
the agreement between DFT and the simulation data is rather good. 
For high polydispersity ($a=26$) the agreement is reasonable only for 
lower values of the Langevin susceptibilities.

\begin{table}
\caption{Size analysis data \cite{ivanov4} of magnetic particles from 
magnetite based ferrofluids at two concentrations.}
\begin{center}
\begin{tabular}{ | c | c | c | c | c | c | c | }
  \hline
 & $\rho$ ($\rm{m^{-3}}$) &  $M_s$ ($\rm{kA/m}$) & $4\pi\chi_L$ & $4\pi\chi_0$ &$\sigma_0 ($\rm{nm}$)$ & $a$ \\
  \hline
ferrofluid I  & 93.8$\times{10^{22}}$ & 72.4& 0.99 & 1.3 &0.39 & 15.3 \\
  \hline
ferrofluid II & 43.8$\times{10^{22}}$ & 87.1& 4.05 & 10.0 &0.97 & 7.54 \\
  \hline
\end{tabular}
\end{center}
\end{table}
On the basis of the experimental data reported in Refs. \cite{pshe1,pshe3} for 
two ferrofluids (containing magnetite particles dissolved in hydrocarbon liquids) 
Ivanov and Kuznetsova \cite{ivanov4} carried out a size analysis of the magnetic 
particles.  Their results are summarized in Table 1. 
They used thermodynamic perturbation theory in order to obtain the corresponding quantities. 
Using the present theory and the parameter sets obtained by Ivanov and Kuznetsova \cite{ivanov4} we 
have calculated the corresponding magnetization curves at room temperature $T=293 K$.

In order to address the parameter sets used in Ref. \cite{ivanov4} the saturation 
magnetizations $M_s={\lim_{_{H\rightarrow\infty}}}{M(H)}$ of the polydisperse 
systems have to be calculated within the framework of the present theory.
Using the asymptotic behavior $\lim_{_{z\rightarrow\infty}}L(z)=1$ of the Langevin function, 
from Eqs. (\ref{eq:magn}) and (\ref{mave}) one obtains
\begin{eqnarray}
\!\!\!\!\!\!\!\!\!\!\!\!
M_s=
\lim_{H\rightarrow\infty}\rho\int_0^{\infty}d{\sigma}p_a(\sigma)m(\sigma)
L\left[{\beta}m(\sigma)\left(H+\frac{(1-q(-\xi))}{\chi_L}M\right)\right]=\nonumber\\
\rho\int_0^{\infty}d{\sigma}p_a(\sigma)m(\sigma)=\rho\langle{m}\rangle_p=
\rho\frac{\pi}{6}M_0\sigma_0^3\prod_{i=1}^3(a+i).
\end{eqnarray}
The comparison between our theoretical predictions and the actual 
experimental data is shown in Figure 5. For ferrofluid I the calculated Langevin 
susceptibility is $4\pi\chi_L\,\,{\simeq}\,\,1$ 
while for ferrofluid II this is $4\pi\chi_L\,\,{\simeq}\,\,4$. 
In Ref. \cite{szalai2} it was found that for \textit{binary mixtures} there is 
quantitative agreement between the results of MSA based DFT and MC 
simulation data only for those Langevin susceptibilities which satisfy the inequality $4\pi\chi_L\,\,{\lesssim}\,\,2.5$.  

\begin{figure}
\begin{center}
\scalebox{0.45}{*\includegraphics{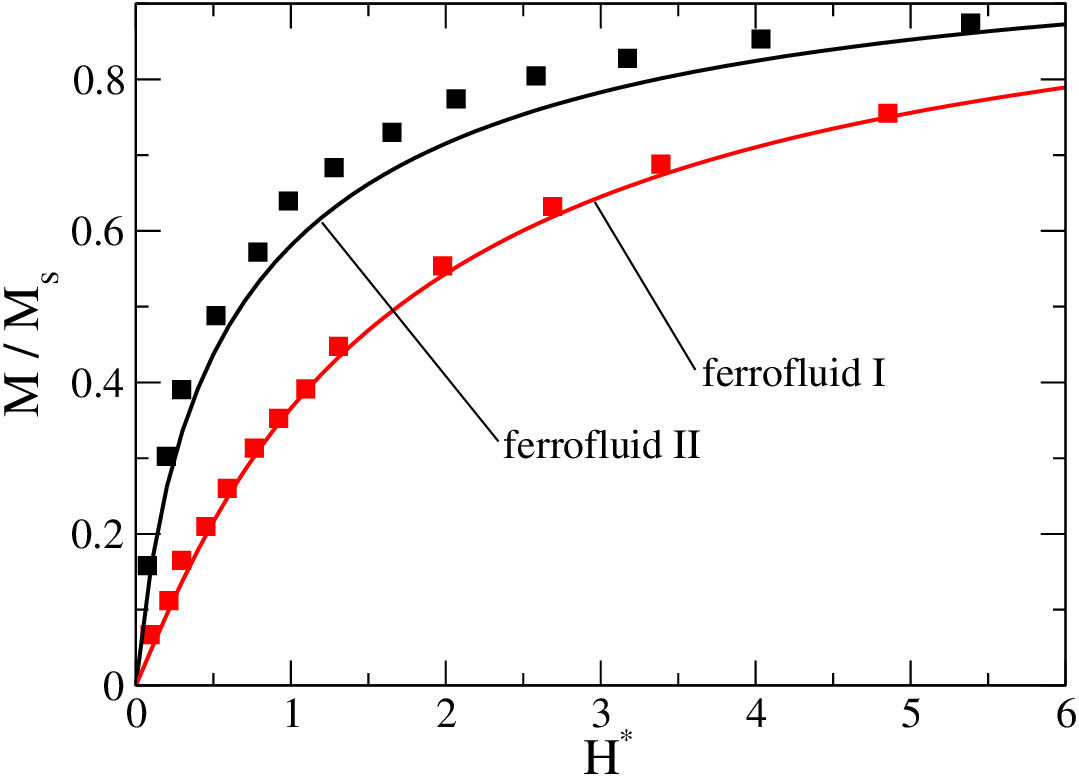}}
\end{center}
\caption{Magnetization curves of two polydisperse magnetite-based ferrofluids 
as function of $H^*=H\sqrt{\langle\sigma^3\rangle_p/(k_BT)}$. 
The theoretical magnetization curves (Eq. (\ref{eq:magn})) are given by full lines. 
The symbols represent the experimental data. For ferrofluid I the parameters of 
the theoretical curve are: $\langle{m^2}\rangle^*=0.76$, $\rho^*=0.29$, and $a=15.3$. 
The corresponding parameters for ferrofluid II are: $\langle{m^2}\rangle^*=2.79$, $\rho^*=0.34$, and $a=7.54$.}
\label{fig5}
\end{figure} 
Here, for the magnetization curves of \textit{polydisperse} systems we have found that one can 
expect satisfactory agreement only for $4\pi\chi_L\,\,{\lesssim}\,\,2$.
This checks with the observation that our theory describes well only 
the experimental data for ferrofluid I.
In Ref. \cite{ivanov3} Ivanov and coworkers compared their MC magnetization data 
for the DHS model as well as their MD magnetization data for the dipolar soft-sphere 
model with the corresponding experimental data of Refs. \cite{pshe2}, \cite{morozov1}, 
and \cite{morozov2}. For both models they found excellent agreement between the simulation 
and the experimental data for $4\pi\chi_L\,\,{\lesssim}\,\,4$ and $a=4.9518$. 
That means that within this range of parameters both the DHS and the dipolar soft-sphere model 
are appropriate to model the interparticle interaction of magnetic grains. 
(This also means that within this parameter range the magnetization data cannot 
discriminate between the DHS and the dipolar soft-sphere model.) 
Therefore the fact that our MSA based DFT describes the field dependence of 
the magnetization only for the Langevin susceptibility values 
$4\pi\chi_L\,\,{\lesssim}\,\,2$ and shape parameters ${a}\gtrsim{13}$ of the gamma 
distribution points towards a restriction on the quantitative reliability of this analytic theory.
\section{Summary}
We have obtained the following results:\\
(1) Based on the MSA theory for multicomponent DHS fluids, an implicit analytical 
expression for the magnetization 
equation of state has been proposed for size polydisperse ferrofluids 
(Eq. (\ref{eq:magn})). The polydispersity 
of the grain diameter is described in terms of the gamma distribution function 
(Eqs. (\ref{eq:dis}) and (\ref{eq:reddis2}) and Fig. 1).\\
(2) We have found that for Langevin susceptibility values $4\pi\chi_L\lesssim{2}$ and shape parameters ${a}\gtrsim{13}$ of the gamma distribution the field dependence of these theoretical magnetization data is in good quantitative agreement 
with corresponding MC simulation data (Figs. 2 and 3).\\
(3) For polydisperse systems we have compared the dependence of the MSA zero-field susceptibility $\chi_0$ (Eq. (\ref{chi0})) on $\chi_L$ (Eq. (\ref{eq:slan})) 
(which can be expressed in terms of a single master curve) with corresponding MC simulation data. There is good agreement for $4\pi\chi_L\lesssim{2}$ and ${a}\gtrsim{26}$ (Fig. 4).\\
(4) Within these parameter ranges for ($\chi_L,a$) we have found also good agreement between our theory and actual experimental data of magnetite-based ferrofluids (Fig. 5 and Table 1).
\section*{Acknowledgments}
I. Szalai and S. Nagy acknowledge the financial support for this work by the Hungarian 
State and the European Union within the TAMOP-4.2.2.A-11/1/ KONV-2012-0071 and 
TAMOP-4.2.2.B-10/1-2010-0025 projects.

\end{document}